\documentclass[aps,pre,twocolumn,showpacs]{revtex4-1}
\usepackage{epsfig,amssymb,amsmath,amsfonts}

\begin{document}
\title{Competition of simple and complex adoption on interdependent networks }
\author{Agnieszka Czaplicka, Raul Toral, Maxi San Miguel}
\affiliation{Instituto de F{\'\i}sica Interdisciplinar y Sistemas Complejos, IFISC (CSIC-UIB), Campus UIB, 07122 Palma de Mallorca, Spain}
\date{\today}

\begin{abstract}
We consider the competition of two mechanisms for adoption processes: a so-called complex threshold dynamics and a simple Susceptible-Infected-Susceptible (SIS) model. Separately, these mechanisms lead, respectively, to first order and continuous transitions between non-adoption and adoption phases. We consider two interconnected layers. While all nodes on the first layer follow the complex adoption process, all nodes on the second layer follow the simple adoption process.  Coupling between the two adoption processes occurs as a result of the inclusion of some additional interconnections between layers. We find that the transition points and also the nature of the transitions are modified in the coupled dynamics. In the complex adoption layer, the critical threshold required for extension of adoption increases with interlayer connectivity whereas in the case of an isolated single network it would decrease with average connectivity. In addition, the transition can become continuous depending on the detailed inter and intralayer connectivities. In the SIS layer, any interlayer connectivity leads to the extension of the adopter phase. Besides, a new transition appears as as sudden drop of the fraction of adopters in the SIS layer. The main numerical findings are described by a mean-field type analytical approach appropriately developed for the threshold-SIS coupled system.
\end{abstract}
\maketitle

\section{Introduction}

Dynamical collective phenomena emerging from interacting units show nontrivial dependencies on the topology and other characteristics of the network of interactions. Examples of this dependency in complex networks occur in synchronization phenomena \cite{Nishikawa2003,Arenas2006,Arenas2008}, in ordering dynamics, where coarsening only occurs below a critical effective dimension\cite{suchecki}, or in the appearance of new types of phase transitions, such as explosive percolation or explosive synchronization \cite{achlioptas,radicchi1,grasberger,gomezgardenes,nagler,warnke}. In these phenomena, an additional important aspect is the multilayering of the network or the existence of underlying interdependent networks \cite{rev1,rev2,rev3,havlin0}. In general, dynamical processes in this type of networks are non-reducible to dynamics in a single effective network \cite{Domenico2014,Diakonova2016} and new forms of phase transitions and changes in the order of the transition have been found in such multilayared/interdependent networks \cite{havlin1,havlin2,havlin3,havlin4,radicchi2,radicchiNature,bianconi}. While most studies of dynamics in complex networks isolate a single dynamical process, many real situations involve the coupling of two processes, a situation most naturally described in the context of multilayer/interdependent networks \cite{havlin0,awareness,awareness2,skardal}. In this paper we consider the situation of interdependent networks where nodes in each layer are distinct entities which follow only one type of contagion mechanism, but due to interconnections are influenced by the nodes from both layers. We address the question of the effect of coupling two dynamical processes featuring, respectively, a discontinuous and a continuous phase transition.

As a specific illustration we focus on contagion processes that describe adoption phenomena. The adoption of an innovation or a new technology can follow from processes of {\slshape simple} or {\slshape complex} contagion \cite{eguiluz,contagion,weng}. In simple contagion a node in the network adopts by interacting with a single neighbor who has already adopted, in the same way than in an infection process. On the contrary, in complex contagion, adoption requires simultaneous exposure to multiple neighboring nodes that have already adopted, so that adoption depends on the global state of the neighborhood. A prototype model for simple contagion is the SIS (Susceptible-Infected-Susceptible) model, known to have a continuous transition to the adoption phase for a critical value of the infection rate \cite{sis,sis2}. Complex contagion is described by a threshold model \cite{Thr} which has a discontinuous transition \cite{watts} to the adoption phase at a critical value of the fraction of neighbors required for individual adoption. A number of recent studies \cite{juancarlos,szymanski,kertesz} consider different aspects of this model, including comparison with available data from online interactions \cite{contagion,may2011,skype,karsai}. While coupling of two simple \cite{sisSerrano,moreno,stanley,shai,jesus,Buono2014,sahneh,Fede2015} or two complex \cite{thrGoh,thrGoh2,mixingThr} contagion processes in multilayared/interdependent networks has been discussed in different situations, our goal here is to consider the coupling of a SIS and a threshold contagion models, each of them running in one of two interdependent network layers. The proposed set-up is motivated by the study of the adoption of a given innovation by two interdependent populations each of which follows a different contagion mechanism, or alternatively the coupled adoption of two different innovations each one associated with a different contagion mechanism. Concerning the fundamental question of the coupling of a continuous and a discontinuous phase transition, we find that not only the transition points and the nature of the transitions can change, but also new transitions appear. These changes turn out to depend on the interlayer connectivity and on the asymmetry between the average intralayer connectivity of the two layers. 

This paper is organized as follows. In section \ref{sec_model} we define the model and give precise dynamical rules for the evolution of the state variables. In section \ref{MF_approach} we introduce a mean-field approximation which allows us to derive evolution equations for the density of adopters in each layer and study the fixed points and their stability, finding the phase diagram in the uncoupled and coupled cases. The more technical details of the derivation of the evolution equations for the density of adopters in the threshold and SIS layers are given, respectively, in appendices A and B. In section \ref{sec_numerical} we show the results of numerical simulations and compare them with the analytical predictions of the previous section. In section \ref{sec_order} we discuss the continuous or discontinuous orders of the different transitions found in the model. Finally, in section \ref{sec_conclusions} we summarize the main conclusions.

\section{Model}
\label{sec_model}
We consider two interdependent \cite{havlin0, havlin1,havlin2,havlin3} layers. Each layer $\ell=1,2$ has $N_\ell$ nodes connected as a random Erd\"os-R\'enyi (ER) topology with average degree $k_\ell$. Additionally, there are $M$ interlinks randomly connecting nodes in both layers. Throughout the paper, we consider equal size layers, $N_1=N_2\equiv N$ and denote by $m=M/N$ the average number of interlinks per node. Each node holds a binary state variable $s_{\ell,i}$, $\ell=1,2,\,i=1,\dots,N$, taken as $s_{\ell,i}=0$ (not adopter, neutral) and $s_{\ell,i}=1$ (adopter). Nodes in the two layers are distinct entities such that each individual is subjected to only one type of contagion mechanism, not to two simultaneously.

In the first layer, $\ell=1$, nodes change their states through a {\slshape complex} adoption process \cite{contagion,eguiluz}, following a variant of the threshold model rules \cite{Thr}: a neutral node switches to the adoption state when the fraction of its adopter neighbors is above a threshold value $\theta$. It is assumed that adoption is irreversible and an adopter can not go back to the neutral state. On a single, uncoupled, network the threshold model displays a discontinuous transition \cite{watts} at a critical value $\theta_c$ which decreases as $1/k_1$. Below the critical threshold $\theta<\theta_c$ all nodes in the system become eventually adopters, while for $\theta>\theta_c$ adoption does not spread and only the initial group of adopters remains.

In the second layer, $\ell=2$, nodes evolve by a {\slshape simple } adoption process following a variant of the SIS dynamics: adoption spreads by pairwise interactions between adopter and non-adopter nodes. The probability that an interaction between a non-adopter and an adopter node leads to adoption in the neutral node is $\lambda$. SIS dynamics does allow adopter nodes to become neutral again. A single, uncoupled network, displays a continuous phase transition at a critical value $\lambda_c$: when $\lambda<\lambda_c$ all nodes end up in the non-adopter state, while for $\lambda>\lambda_c$ there is a non-vanishing fraction of nodes that become adopters. For our particular rules, we find $\lambda_c=1/k_2$, see section ~\ref{section:fixed_points}.

\begin{figure}
\centerline{\psfig{file=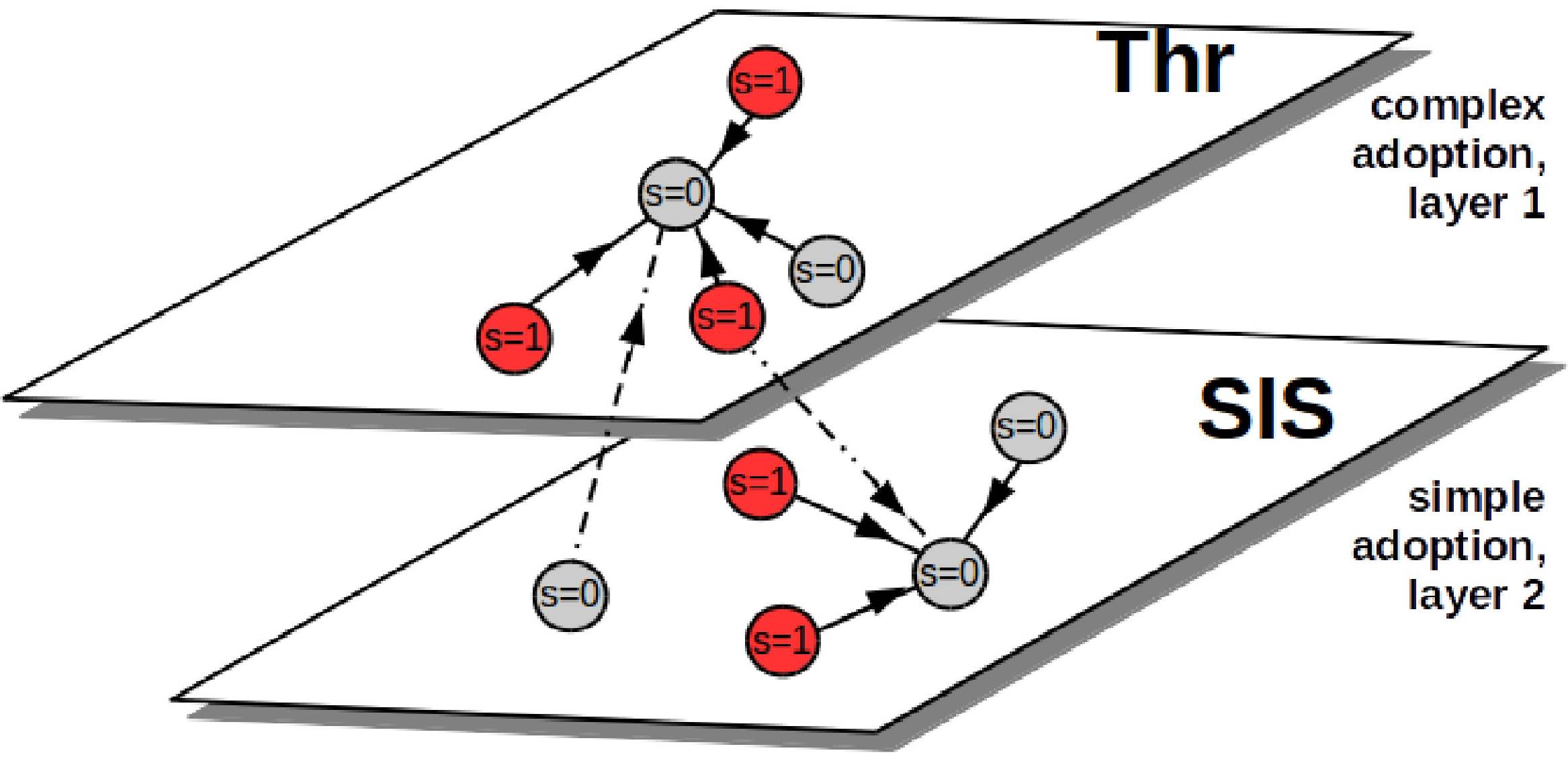,width=0.99\columnwidth}}
\caption{Schematic representation of the model's dynamics in two layers forming an interdependent network in which a node can interact with neighbors from both layers. In the upper layer (Thr) nodes change their states through a complex adoption process following a threshold model rules. In the lower layer (SIS) nodes evolve by a simple adoption process following the SIS type of dynamics.}
\label{Fig:2Lschem}
\end{figure}

When the networks are coupled, $m>0$, the detailed dynamics is as follows: we start at $t=0$ with a small seed of adopters in the first layer (one randomly chosen node and all its neighbours in the first layer) and a single adopter in the second layer. At successive time steps $t>0$ we choose randomly one layer $\ell$ and one node $i$ from this layer, and its state is updated according to the following rules (see Fig.~\ref{Fig:2Lschem} for a schematic representation of these rules):\\
$\bullet$ If the node belongs to the first layer and
\begin{itemize}
\item[-] $s_{1,i}(t)=0$, then $s_{1,i}(t+1)=1$ if at least a fraction $\theta$ of its neighbours (from both layers) are adopters,
\item[-] $s_{1,i}(t)=1$, nothing happens. Adopters cannot become neutral.
\end{itemize}
$\bullet$ If the node belongs to the second layer and
 \begin{itemize}
\item[-] $s_{2,i}(t)=0$, then all neighbors (from both layers) of the node are visited sequentially. Adoption from any contact arises with probability $\lambda$.
\item[-] $s_{2,i}(t)=1$, then it goes back to neutral state.
\end{itemize}

\section{Mean-field approach to system's dynamics}\label{MF_approach}
\subsection{Threshold layer}
Threshold dynamics on a single network had been previously described analytically (e.g. see \cite{watts,thrGoh,gleeson2}). In these treatments the authors used assumptions that are not valid in our particular case because of the coupling between two different types of dynamics. For instance, we can not assume in the SIS layer that a node can change its state at most once during the evolution, or it is not convenient to treat the network as tree-like for large $m$ values. Therefore we develop in this paper an approach based on a mean-field type approximation in which local fractions are replaced by global averages. 
As shown in Appendix A, under this approximation the evolution equation for the fraction of adopters in threshold layer is
\begin{equation}
\beta\dfrac{d\langle s_1(t)\rangle}{dt}= \left(1-\langle s_1(t)\rangle\right)\text{Prob}\left[\langle s_{1}(t)\rangle_\textrm{n}\ge \theta\right],
\label{g1}
\end{equation}
with $\beta=\dfrac{N_1}{N_1+N_2}$ and $\langle s_{1}\rangle_\textrm{n}$ is the average fraction of neighbours which are adopters. 

Although it is possible to relax this condition, to proceed further we assume that each site $i$ in the first layer has exactly $k_1$ neighbors in layer 1 and $m$ neighbors in layer 2. Within the mean-field approximation it can be assumed that the average fraction of neighbours which are adopters is a weighted average of the average number of neighbors in each layer:
\begin{equation}
\langle s_{1}\rangle_\textrm{n} = \frac{k_1 \langle s_1 \rangle + m \langle s_2 \rangle }{k_1 + m}.
\label{s1k1m_appendix}
\end{equation}

The probability to find exactly $j$ adopters amongst the $k_1+m$ neighbors, given that the fraction of adopters is $\langle s_{1}\rangle_\textrm{n}$, follows a binomial distribution
\begin{equation}
B\left(\langle s_{1}\rangle_n, j\right)\equiv{{k_1 + m} \choose {j}} (\langle s_{1}\rangle_n)^{j}(1-\langle s_{1}\rangle_n )^{k_1 + m -j}.
\label{bxy_appendix}
\end{equation}
To find the probability that a neutral node becomes adopter, we sum all cases when $j\geq\lfloor\left( k_1 + m \right)\theta\rfloor$, where $\lfloor x\rfloor$ denotes the largest integer not greater than $x$. This yields
\begin{equation}
\text{Prob}\left[\langle s_{1}(t)\rangle_\textrm{n}\ge \theta\right]=\sum_{j=\lfloor\left(k_1 + m \right)\theta\rfloor}^{k_1 + m}B(\langle s_{1}\rangle_\textrm{n}, j).
\label{g1_appendix}
\end{equation}
The binomial distribution Eq.~(\ref{bxy_appendix}) can be approximated by a Gaussian distribution with mean $(k_1 + m)\langle s_{1}\rangle_\textrm{n}$ and variance $(k_1 + m)\langle s_{1}\rangle_\textrm{n}(1-\langle s_{1}\rangle_\textrm{n})$, leading to
\begin{widetext}
\begin{equation}
\text{Prob}\left[\langle s_{1}(t)\rangle_\textrm{n}\ge \theta\right]\approx \frac{1}{2}\textrm{erfc}\left[\frac{\lfloor(k_1+m)\theta)\rfloor-1/2-(k_1+m)\langle s_{1}\rangle_\textrm{n})}{\sqrt{2(k_1+m)\langle s_{1}\rangle_\textrm{n}(1-\langle s_{1}\rangle_\textrm{n})}}\right],
\label{g1limit_appendix}
\end{equation}
\end{widetext}
where $\textrm{erfc}[x]$ is the complementary error function. This turns out to be a good numerical approximation for not too large values of $k_1+m$ (an error small than $0.1$ in the whole range of $\theta$ for $k_1+m=10$). In the limit of large $k_1+m$ this can be further approximated by
\begin{equation}
\text{Prob}\left[\langle s_{1}(t)\rangle_\textrm{n}\ge \theta\right]\approx \frac{1}{2}\textrm{erfc}\left[\frac{\theta-\langle s_{1}\rangle_\textrm{n}}{\sqrt{\frac{2}{k_1+m}\langle s_{1}\rangle_\textrm{n}(1-\langle s_{1}\rangle_\textrm{n})}}\right].
\label{g1limit2_appendix}
\end{equation}
We still need a final ingredient to derive the evolution equation for $\langle s_1(t)\rangle$. As we are studying the evolution in a finite system, if the probability $\text{Prob}\left[\langle s_{1}(t)\rangle_\textrm{n}\ge \theta\right]$ is smaller that $1/N_1$, it means effectively that the condition can not be reached in the finite system. Therefore, we introduce the function
\begin{equation}
G(x)=\begin{cases}\text{Prob}\left[ x\ge \theta\right], & \mbox{if} \; \text{Prob}\left[ x\ge \theta\right]\ge\frac{1}{N_1},\\
0, &\mbox{if} \; \text{Prob}\left[x\ge \theta\right]<\frac{1}{N_1},\end{cases}
\end{equation}
and then write the evolution equation as:
\begin{equation}
\beta \dfrac{d\langle s_1(t)\rangle}{dt}=\left(1-\langle s_1\rangle\right)G\left(\frac{k_1 \langle s_1 \rangle + m \langle s_2 \rangle }{k_1 + m}\right),
\label{thr4}
\end{equation}
where we have used Eq.~(\ref{s1k1m_appendix}). This is the final mean-field equation for the evolution of the density of adopters in the threshold layer.

\subsection{SIS layer}
In the usual Susceptible-Infected-Susceptible (SIS) model of infection \cite{sis,sis2}, the evolution rules are that a randomly chosen agent which happens to be adopter (``infected'' state) can either infect one of its neighbours with a given probability or can go back to the neutral (``susceptible'' state); if the randomly chosen agent is already in the neutral state, nothing happens. Here we use a slightly modified version of the SIS rules that we believe are more appropriate to model adoption processes. In this version it is the neutral node the one adopting from its adjacent neighbors adopters. The way of interaction remains the same, pairwise interactions between nodes are considered, but the direction of interaction has been changed from outgoing (original SIS for infection) to ingoing (adoption process). In this way we keep the symmetry in the interaction between the threshold and the SIS layers and we implement two-way influence between complex and simple adoption layers. By a similar reasoning to the one developed before for the threshold layer, we can derive the mean-field equation for the fraction of adopters in the SIS layer (details of the derivation are given in Appendix B)
 \begin{equation}
(1-\beta)\dfrac{d\langle s_2(t)\rangle}{dt}= -\langle s_2\rangle+\left(1-\langle s_2\rangle\right)\left(1-e^{-\lambda\left(k_2 \langle s_2 \rangle + m \langle s_1\rangle\right)}\right).
\label{sis4}
\end{equation}
Equations (\ref{thr4}) and (\ref{sis4}) are the starting point of our analytical treatment. In the next section we discuss the fixed points and their stability. For the sake of brevity we adopt henceforth the notation $x_1\equiv\langle s_1\rangle$, $x_2\equiv\langle s_2\rangle$.

\subsection{Fixed points}
\subsubsection{Independent layers}\label{section:fixed_points}
We first analyze the fixed points in the absence of coupling between the layers, $m=0$.

\begin{figure*}
\centerline{\psfig{file=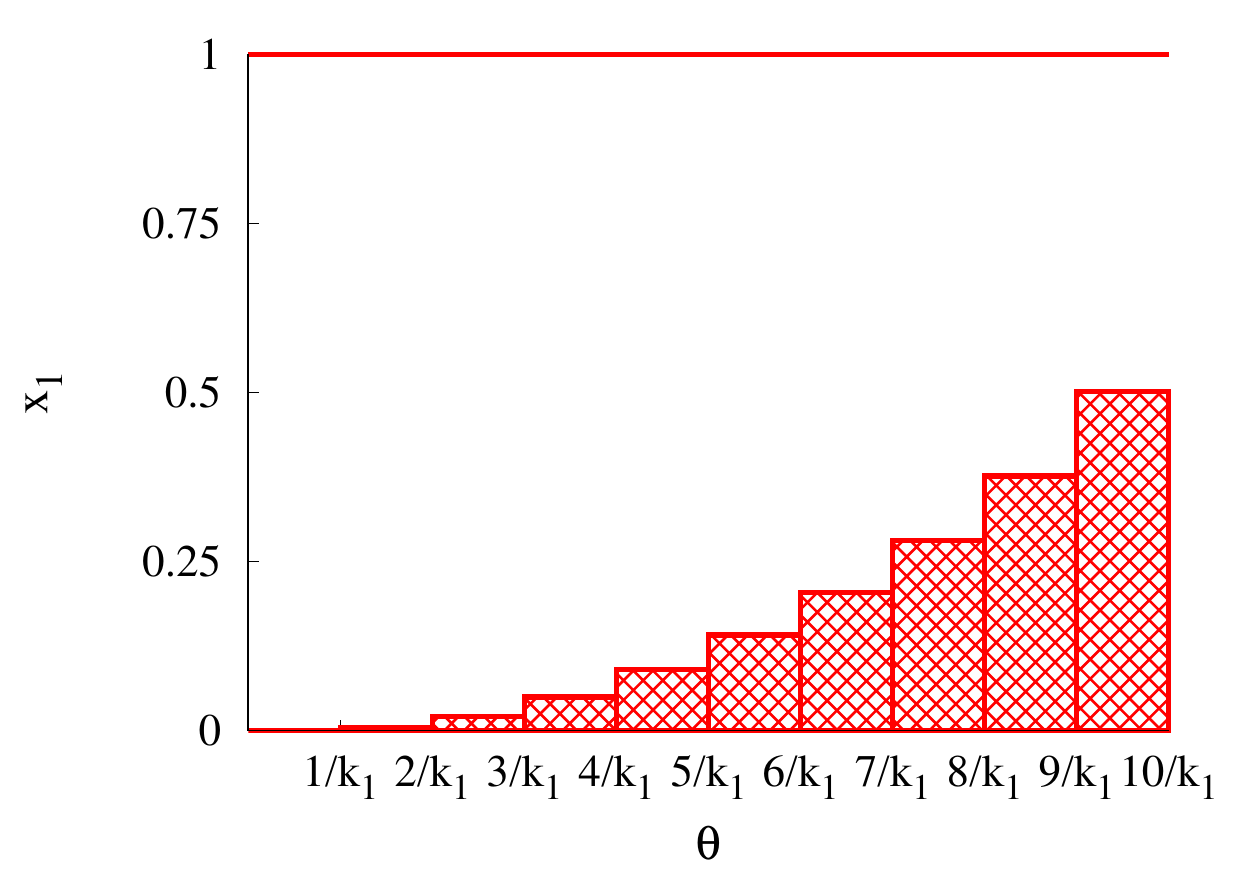,width=0.99\columnwidth}\psfig{file=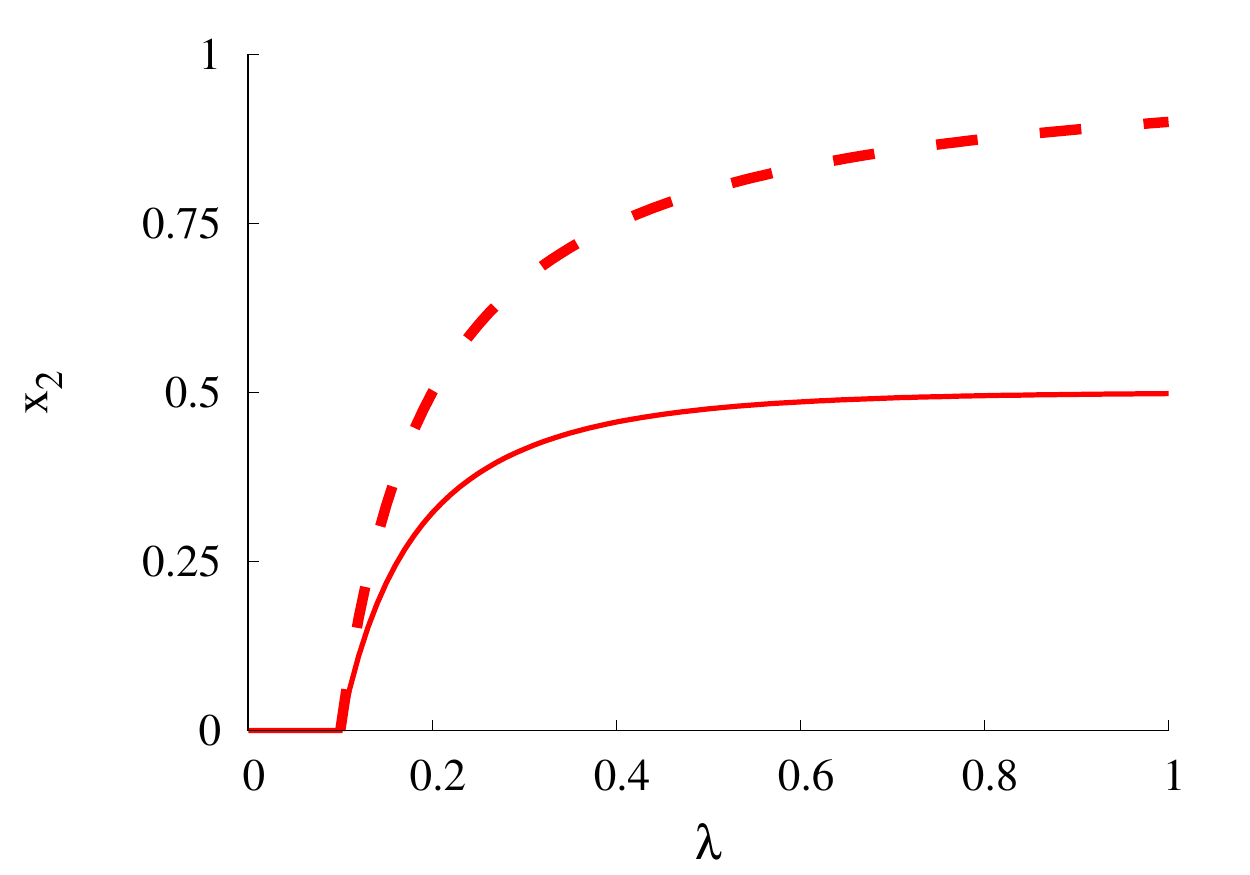,width=0.99\columnwidth}}
\caption{Phase diagram in the independent layer case $m=0$.  Left-side panel shows the case of the threshold layer for $k_1=10$, $N_1=1000$. Here the dashed area and the line $x_1=1$ is the set of stable fixed points. The right panel displays the steady state fraction of adopters for the $SIS$ model $k_2=10$, $N_2=1000$. Dashed line presents results for $usual$ model of infection, solid line presents our model of simple adoption. }
\label{Fig:phased_x1_x2}
\end{figure*}

For the first, threshold, layer, Eq.~(\ref{thr4}) has always the stable fixed point $x_1^*=1$. New fixed points appear as solution of the equation
\begin{equation}
G\left(x_1^*\right)=0.
\end{equation}
If $\lfloor k_1 \theta\rfloor=0$, the sum in Eq.~(\ref{g1}) is always equal to $1$ and there are no new fixed points. They appear when $\lfloor k_1 \theta\rfloor=1$, or $\theta=\Delta \theta\equiv1/k_1$, as now the sum Eq.~(\ref{g1}) misses the term with $j=0$ and hence it is equal to $1-(1-x_1^*)^{k_1}$. According to its definition, $G(x_1^*)=$ 0 if $1-(1-x_1^*)^{k_1}<1/N_1$ or $\displaystyle x_1^*<1-\left(1-\dfrac{1}{N_1}\right)^{1/k_1}\approx (N_1k_1)^{-1}$, for large $N_1$. When $\theta=2\Delta \theta$ the sum Eq.~(\ref{g1}) misses two terms and the interval of fixed points is given by the condition $1-(1-x_1^*)^{k_1}-k_1x_1(1-x_1^*)^{k_1-1}<1/N_1$, or $x_1^*\lesssim\sqrt{\dfrac{2}{N_1k_1(k_1-1)}}$, for large $N_1$. The appearance of an enlarged interval of fixed points continues until $\theta=1$, where the only term in the sum in Eq.~(\ref{g1}) is $(x_1^*)^{k_1}$ and the interval of fixed points is $x_1^*< N_1^{-1/k_1}<1$. The complete phase diagram for the uncoupled threshold layer is plotted in Fig.~\ref{Fig:phased_x1_x2}, left panel. If the initial condition $x_1(0)$ falls inside the shaded area, then it remains there. If, otherwise, the initial condition is outside the shaded area, then the dynamics leads to the only stable stationary solution, $x_1^*=1$, corresponding to global adoption. As the initial condition can not be smaller that $1/N_1$ (one single adopter) one must consider adoption possible only whenever $x_1^*>1/N_1$. In our particular version of the threshold model for a wide range of reasonable values of $N_1$ and $k_1$ this occurs for $\theta=\theta_c=2/k_1$. A more precise treatment was presented in reference \cite{watts}. It was found there that the condition for $\theta_c$ for threshold dynamics on ER graphs is 
\begin{equation}
 k_1 Q(K_*-1,k_1)=1,
\end{equation}
where $Q(a,x)$ is the incomplete gamma function and $K_*=\lfloor 1/\theta_c\rfloor$. When the initial group of adopters is sufficiently small (three orders of magnitude less than the number of nodes) an approximation for $\theta_c\approx 1/k_1$ can be used. Both in this more detailed calculation, and in our simple treatment, it is found that $\theta_c$ varies as the inverse of the number of neighbors $k_1$.

For the second, SIS, layer, Eq.~(\ref{sis4}) always possesses the solution $x_2^*=0$. This is stable up to $\lambda \le\lambda_c\equiv1/k_2$. A transcritical bifurcation leads to a new, stable, fixed point $x_2^*\in[0,1/2]$ for $\lambda>1/k_2$ appearing as a solution of
\begin{equation}
e^{-\lambda k_2 x_2^*}=\frac{1-2x_2^*}{1-x_2^*}.
\label{Eq:x2}
\end{equation}
The complete phase diagram for the uncoupled threshold layer is plotted in Fig.~\ref{Fig:phased_x1_x2}, right panel. For $\lambda \le \lambda_c$, and independently on the initial condition, the system tends to the only stable fixed point $x_2^*=0$. For $\lambda > \lambda_c$ the dynamics leads to the stable solution $x_2^*>0$ corresponding to partial adoption. Note that the fixed point satisfies $x_2^*\le 1/2$ and hence no more than half of the agents become adopters. In the usual SIS model \cite{sis,sis2} considered on single network the critical value of adoption probability is the same as in our case, i.e. $\lambda_c=1/k_2$ and above this value the fraction of adopters grows as $x_2^*=1-\frac{1}{\lambda k_2}$ (see the dashed line in the right panel of Fig.~\ref{Fig:phased_x1_x2}).

\subsubsection{Coupled layers}
We now consider the case $m>0$.

$x_1^*=1$ is still a fixed point for Eq.~(\ref{thr4}). The corresponding solution $x_2^*$ is obtained from
Eq.~(\ref{sis4})
\begin{equation}
e^{-\lambda k_2 x_2^*}=e^{\lambda m}\frac{1-2x_2^*}{1-x_2^*}.
\end{equation}
It is easy to show graphically that this solution exists for all values of $\lambda$ and it belongs to the interval $x_2^*\in[0,1/2]$, although in general $x_2^*$ has to be found numerically as a a function of $\lambda k_2$ and $\lambda m$. The stability of the fixed point $(x_1,x_2)=(1,x_2^*)$ is analyzed by means of the eigenvalues of the matrix of first derivatives:
\begin{equation}
\begin{pmatrix}\dfrac{\partial \dot x_1}{\partial x_1}& \dfrac{\partial \dot x_1}{\partial x_2}\\ & \\ \dfrac{\partial \dot x_2}{\partial x_1}&\dfrac{\partial \dot x_2}{\partial x_2}\end{pmatrix}
\end{equation}
evaluated at $(1,x_2^*)$. The two eigenvalues $\mu_{1,2}$ are
\begin{eqnarray}
\mu_1&=&-\beta^{-1}G(1,x_2^*),\\
\mu_2&=&\dfrac{-1+k_2\lambda (1-x_2^*)(1-2x_2^*)}{(1-\beta)(1-x_2^*)}.
\end{eqnarray}
While it is clear that $\mu_1<0$, a graphical analysis shows that the second eigenvalue $\mu_2$ is always negative as well and the fixed point $(1,x_2^*)$ is, hence, stable.

Other fixed points might appear as simultaneous solutions of the equations:
\begin{eqnarray}
G\left(\frac{k_1x_1+ m x_2 }{k_1 + m}\right)&=&0,\label{Geq0}\\
e^{-\lambda (k_2 x_2+m x_1)}&=&\frac{1-2x_2}{1-x_2},\label{x1fromx2}
\end{eqnarray}
satisfying the conditions $x_1\in[0,1],x_2\in[0,1]$.

\begin{figure*}
\centerline{\psfig{file=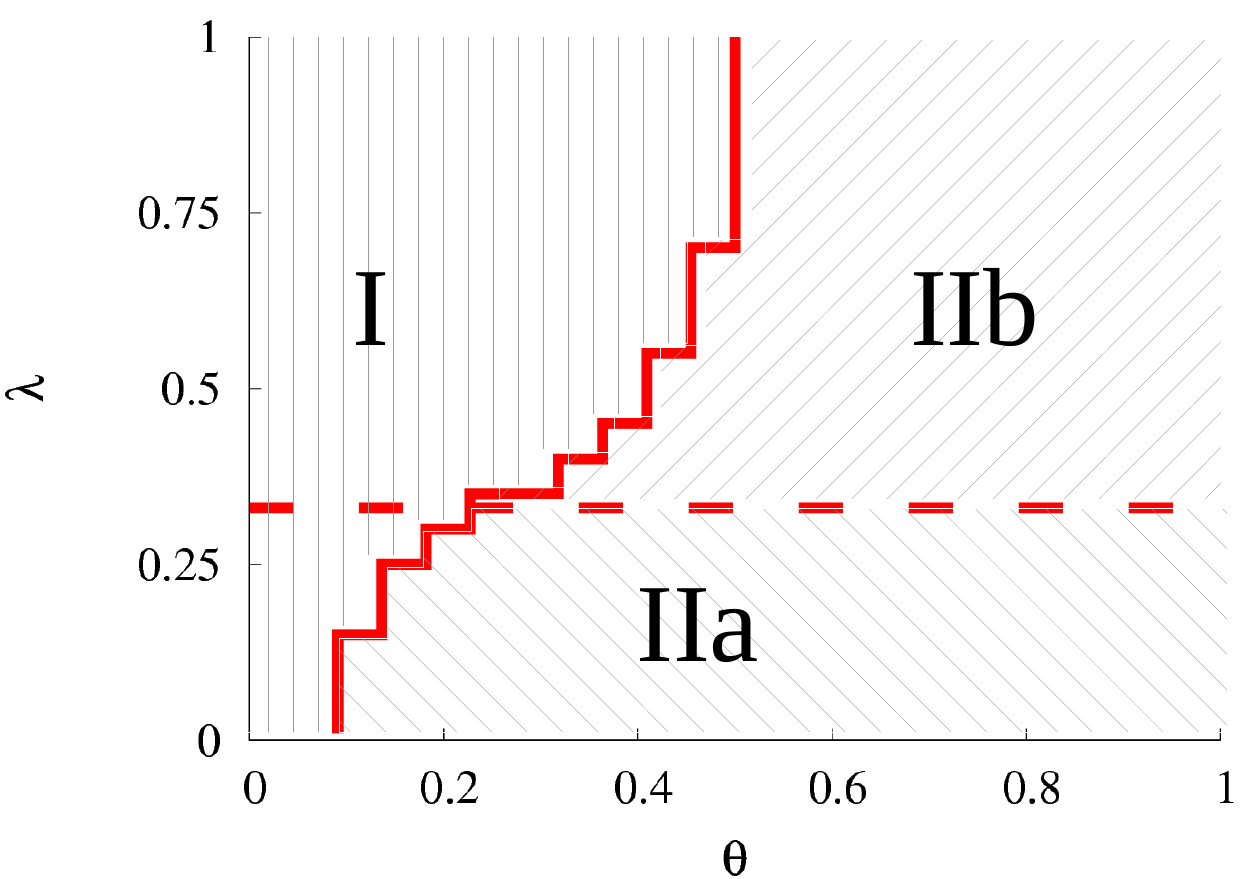,width=0.99\columnwidth}\psfig{file=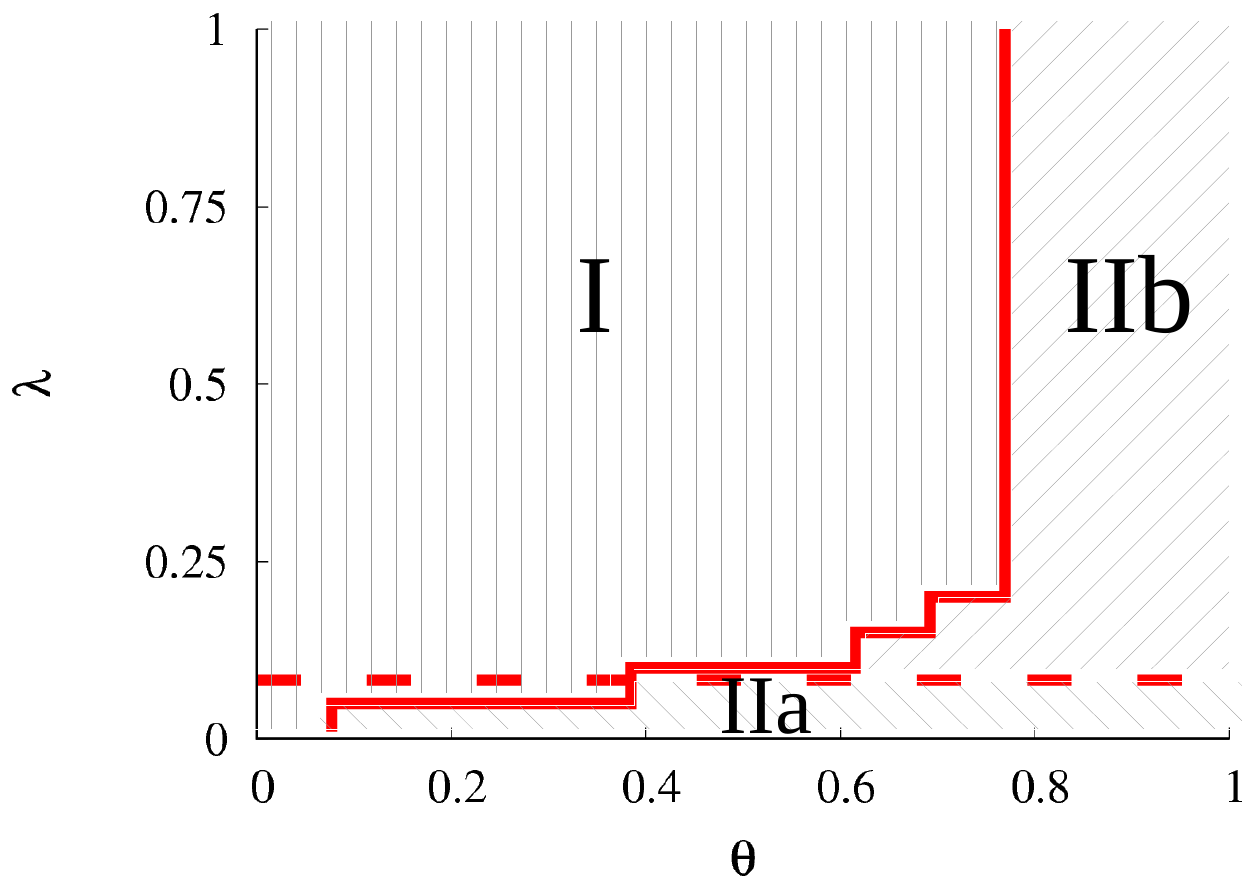,width=0.99\columnwidth}}
\caption{Phase diagram in the coupled layer cases $k_1=12$, $k_2=3$ (left panel) and $k_1=3$, $k_2=12$ (right panel). In both cases it is $m=10$, $N_1=N_2=1000$.}
\label{Fig:phased_x1_x2_m}
\end{figure*}

Replacing $x_1$ from the second equation into the first we arrive at the single condition
\begin{equation}
G\left(\frac{\dfrac{k_1}{\lambda}\log\left(\dfrac{1-x_2}{1-2x_2}\right)+(m^2-k_1k_2)x_2}{m(k_1+m)}\right)=0.
\end{equation}
An analysis similar to the one performed in the case $m=0$ shows that this equation is satisfied in a range of values $x_2\in[0,x_2^*]$ only when $\theta>\Delta \theta\equiv1/(k_1+m)$ (otherwise the function $G$ is identically equal to $1$). However, it might occur that $x_2^*$ is such that the corresponding value for $x_1^*$ obtained from Eq.(\ref{x1fromx2}) does not belong to the interval $[0,1]$. One must then increase $\theta$ until the condition $x_1^*\in[0,1]$ is satisfied. Furthermore, as we discussed above, we must require that $x_1^*$ is greater that $1/N_1$. When these conditions are met, besides finding the fixed points $x_1^*,x_2^*$, one finds the critical value $\theta_c$ as a function of $\lambda$ and the other parameters, $k_1,k_2,m,N_1$ of the model. The resulting line $\theta_c(\lambda)$ is plotted as a solid line in Fig.~\ref{Fig:phased_x1_x2_m} and it has been plotted on top of the numerical results in Fig.~\ref{Fig:phased}. 

To prove that $x_2^*$ is different from zero whenever $\lambda>0$, we expand Eq.(\ref{x1fromx2}) around $\lambda=0$ and replace $x_1^*=1$ to obtain $x_2^*=m\lambda+O(\lambda^2)$. As a way of example, the whole dependence of $x_2^*(m)$ displayed in Fig.~\ref{Fig:s2m} is obtained by a numerical solution of Eq.(\ref{x1fromx2}) fixing $\lambda=0.01$, $x_1^*=1$.

When analyzing the values of $x_1^*$ and $x_2^*$ corresponding to the fixed points, it turns out that if $\lambda<1/k_2$ both values $x_1^*$ and $x_2^*$ are close to zero, but when $\lambda>1/k_2$, $x_1^*$ is still close to zero, but $x_2^*$ takes a value larger than the solution of Eq.(\ref{Eq:x2}). This means that the number of adopters in the SIS layer is always larger in the coupled case than in the uncoupled layers. 

In summary, for $m>0$, the structure of the fixed points of Eqs.(\ref{thr4},\ref{sis4}) is as follows:
\begin{description}
\item[\bf{I}] $x_1^*=1$, $x_2^*\approx 0.5$, for $\theta\leq \theta_c$.
\item[\bf{IIa}] $x_1^*\gtrsim 0$, $x_2^*\gtrsim 0$, for $\theta>\theta_c$ and $\lambda<1/k_2$,
\item[\bf{IIb}] $x_1^*\gtrsim 0$, $x_2^*> 0$, for $\theta>\theta_c$ and $\lambda > 1/k_2$.
\end{description}
These three regimes have been identified in Fig.~\ref{Fig:phased_x1_x2_m} for a particular value of the system parameters. In the next sections we will compare the results of our analytical treatment with those obtained in computer simulations.

\section{Numerical Results compared with analytical findings}	
\label{sec_numerical}
\begin{figure}
\centerline{\psfig{file=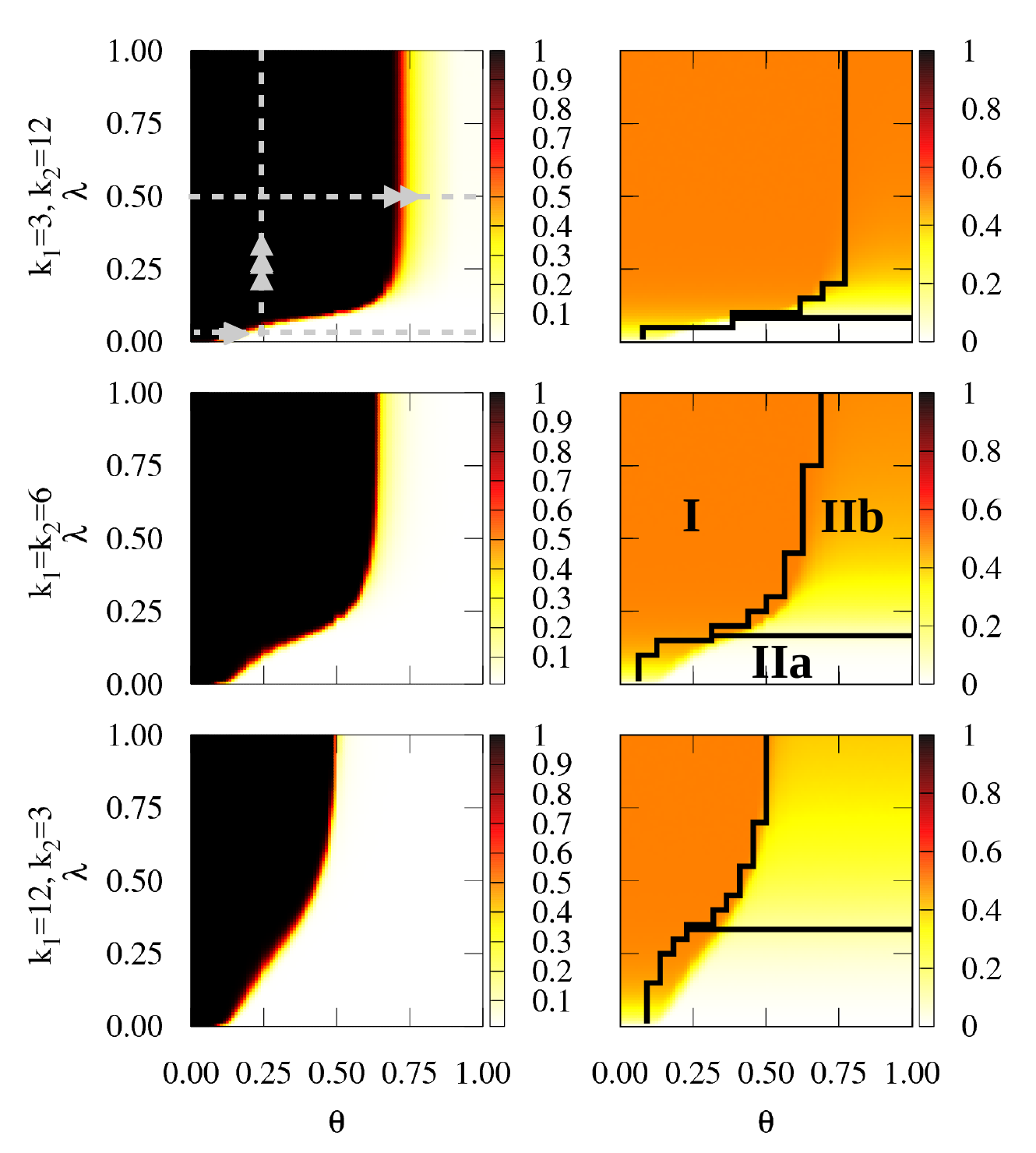,width=0.99\columnwidth}}
\caption{(Color online) Fractions of adopters  in threshold layer $\left<s_1\right>$ (left column), and in SIS layer $\left<s_2\right>$ (right column) as a function of $\theta$ and $\lambda$ for $m=10$ and $k_1=3$ and $k_2=12$ (top panel), $k_1=k_2=6$ (middle panel), $k_1=12$ and $k_2=3$ (bottom panel). Results come after computer simulations performed for a network of $N_1=N_2=1000$ nodes, where both layers are ER random networks, and are averaged over $500$ realizations of networks and $500$ realizations of the dynamics for each network configuration. The continuous lines displayed in the SIS column come from the full stability analysis of the mean-field dynamical equations; the horizontal line is $\lambda=1/k_2$. The arrows indicate transitions between the regions that will be discussed in other figures.}
\label{Fig:phased}
\end{figure}

We have performed numerical simulations of the dynamical rules of the model. We run the dynamics until the steady state is reached (absorbing state in threshold layer but still active in SIS) and then measure the fraction of adopters in each layer $s_\ell=\frac{1}{N_\ell}\sum_{i=1}^{N_\ell}s_{\ell,i}$ and its average $\langle s_\ell\rangle$ over many realizations and network configurations. The results of numerical simulations shown in Fig.~\ref{Fig:phased} evidence that the main effect of the number of interconnections $m$ is to facilitate adoption in the coupled layer system and that there is great correlation between the adoption areas in both layers. 

Our numerical findings are generally well described by the mean-field type analysis described in Section \ref{MF_approach}. The analytical approach is able to predict the main trends observed in Fig.~\ref{Fig:phased}, namely the splitting of the parameter space in distinct regions: region {\bfseries I} characterized by the values $\langle s_1\rangle \approx 1$, $\langle s_2\rangle \approx 0.5$; and region {\bfseries II} characterized by a low value $\langle s_1\rangle \gtrsim 0$, further splitted in regions {\bfseries IIa}: very small $\langle s_2\rangle \gtrsim 0$, and {\bfseries IIb}: larger but still small $\langle s_2\rangle > 0$. The border between regions {\bfseries IIa} and {\bfseries IIb} corresponds to $\lambda=1/k_2$, the critical value in absence of coupling between the layers (see Section \ref{section:fixed_points}). We now describe in details the main features of the different transitions that occur between these regions and whose exact nature depend on the intralayer $(k_1,k_2)$ and interlayer $m$ connectivities.

\begin{figure}[ht]
\psfig{file=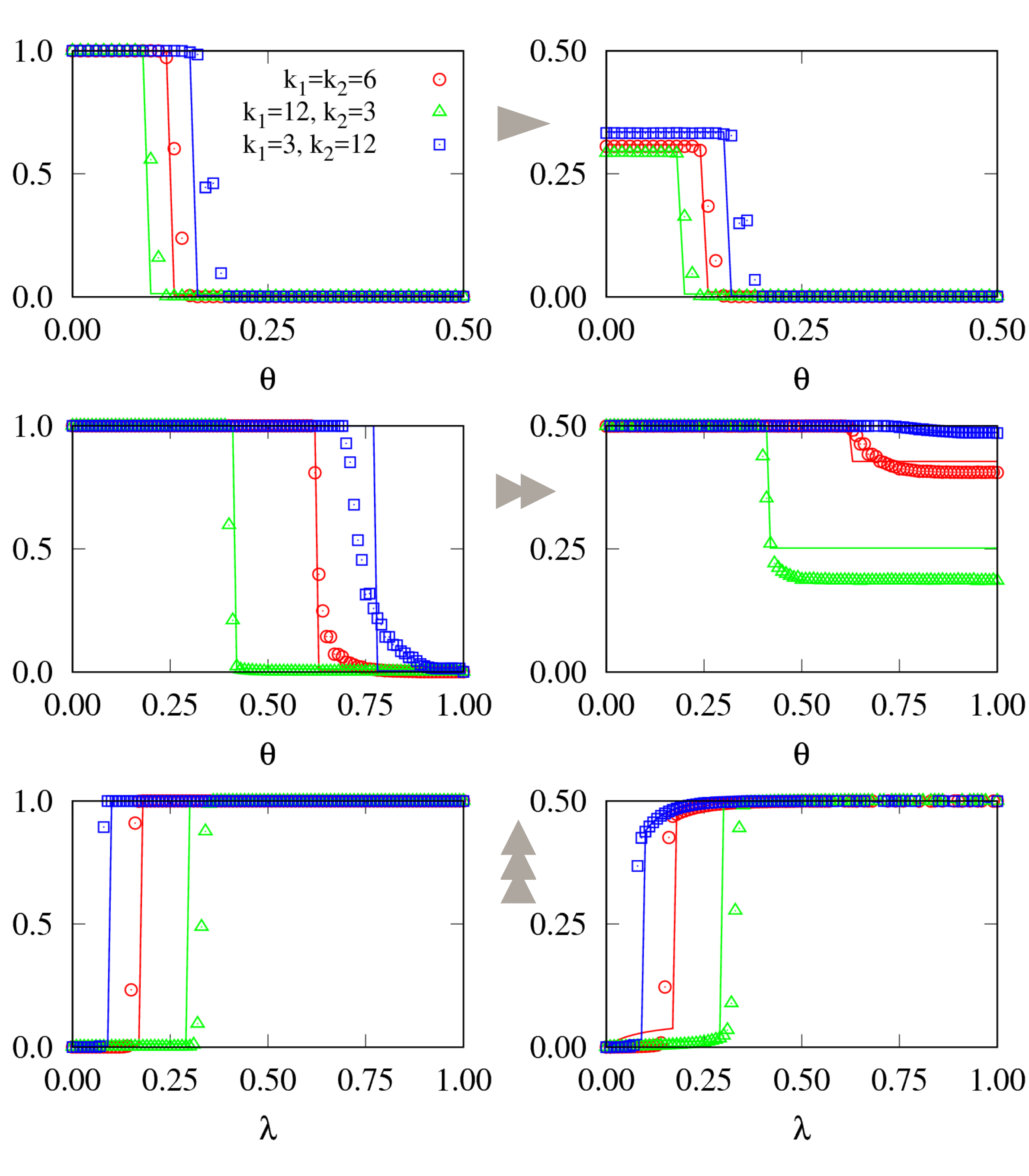,width=0.99\columnwidth, keepaspectratio = true}
\caption{(Color online) Fractions of adopters in threshold layer (left column) and in SIS layer (right column) as a function of $\theta$ (resp. $\lambda$) when $\lambda$ (resp. $\theta$) is fixed and for $m=10$. Top panel shows results when $\lambda=0.05$, middle panel when $\lambda=0.5$, and bottom panel when $\theta=0.25$. Red circles correspond to the case $k_1 = k_2=6$, green triangles to $k_1=12 > k_2=3$, and blue squares to $k_1 =3< k_2=12$. Symbols present the results from computer simulations and solid lines with corresponding colors stand for analytical solutions.} 
\label{Fig:s1s2}
\end{figure}

We first focus on the variation of the adoption region in the threshold layer. As observed in Fig.~\ref{Fig:phased} for a particular value $m=10$ (although results are qualitatively the same for other $m$ values), in the coupled layer system $\langle s_1\rangle$ still experiments a transition from the adopter {\bfseries I} to the neutral {\bfseries II } phase. Evidence of this transition is provided in the left panels of Fig.~\ref{Fig:s1s2} where we plot, for $m=10$ and different values of the intralayer connectivities $(k_1,k_2)$, the variation of $\langle s_1\rangle$ as we cross (i) from region {\bfseries I } to region {\bfseries IIa} (left top panel) varying $\theta$, (ii) from region {\bfseries I } to region {\bfseries IIb} varying $\theta$ (left middle panel), and (iii) from region {\bfseries II } to region {\bfseries I} varying $\lambda$ (left bottom panel). 

For fixed $\lambda$ the transition between regions {\bfseries I } and {\bfseries II } occurs at a critical threshold $\theta_c(\lambda,m)$ that varies with $m$ and $\lambda$ in an intricate way. As $\lambda$ increases towards $\lambda=1$, $\theta_c$ tends to a constant value that depends both on the intralayer ($k_1,\,k_2$) and interlayer ($m$) connectivities. On the one hand, $\theta_c(\lambda=0,m>0)$ is necessarily smaller than the critical value $\theta_c(m=0)$ for the uncoupled case: neighbors in the second layer can not become adopters, hence decreasing the fraction of adopter neighbors of a node in the first layer and making adoption more difficult. On the other hand, the increase of $\theta_c(\lambda>0,m)$ with $m$ seems to go against intuition, since in a single layer $\theta_c$ varies as the inverse of the number of neighbors, which now is $k_1+m$. However, the number of adopters in the second layer can increase due to its own dynamics and, through the interlayer connections, favor the spread in the first layer. The combined effect leads to an increase in the critical threshold value $\theta_c$ for $\lambda,m>0$. 

\begin{figure}[ht]
\psfig{file=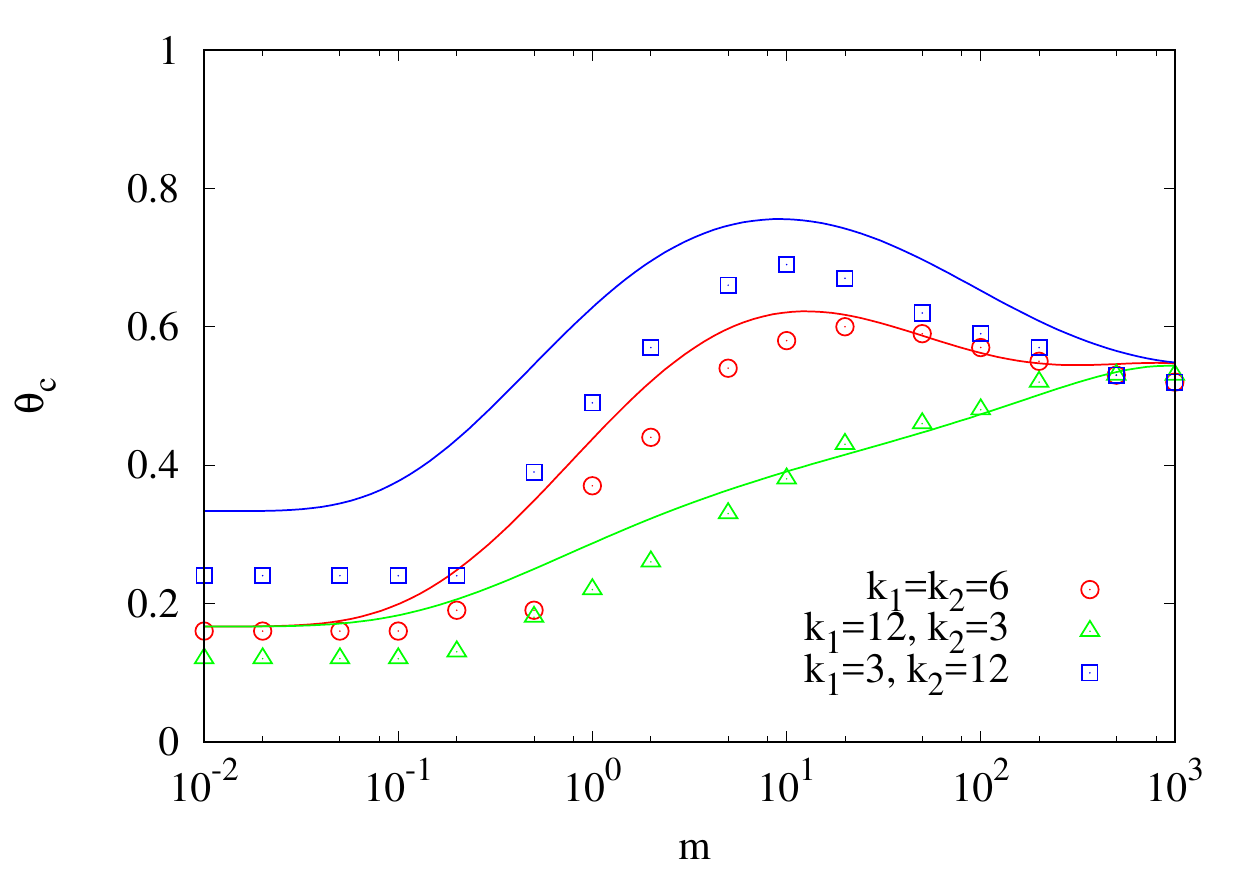,width=0.99\columnwidth, keepaspectratio = true}
\caption{(Color online) Critical value of the threshold $\theta_c$ as a function of the interconnectivity $m$ for $\lambda=0.5$. Same symbols and line meanings than in figure \ref{Fig:s1s2}.}
\label{Fig:thetac}
\end{figure}

We plot in Fig.~\ref{Fig:thetac} the transition value $\theta_c$ as a function of $m$ for a particular value $\lambda=0.5$ (again, similar results are obtained for other values of $\lambda$). There appear to be clear differences between the cases $k_1 > k_2$ and $k_1 \leq k_2$. When $k_1>k_2$, the critical threshold changes visibly only when $m>k_1$, i.e. when the number of interlayer connections overcomes the number of connections inside the threshold layer. It then grows steadily towards the limiting value $\theta_c=0.5$. When $k_1 \leq k_2$, $\theta_c$ increases with the number of interconnections up to a maximum at $m\approx 10$ and then decreases until the limiting value of $\theta_c=0.5$. It is interesting to note that in a single network it is not possible to exceed the value $\theta_c=0.5$ \cite{morris}, whereas at least four coupled networks are needed to observe global adoption above $\theta=0.5$ \cite{thrGoh}. We find that this limit is overcome for just two layers which couple simple and complex contagion processes.

On the SIS layer the most remarkable feature appearing when $m>0$ is the disappearance of the neutral phase: for fixed $\theta$ and $m>0$, $\langle s_2\rangle$ is always larger than zero for $\lambda>0$, or, in other words $\lambda_c(m>0)=0$. This occurs as there are always adopters in the first layer (at least the initial group of adopters remain) and adoption always spreads from them to the second layer due to the interlayer connections. The second noticeable effect of the interlayer connectivity on the SIS layer is the appearance of a new transition when crossing from region {\bfseries I} to region {\bfseries II}. This shows up as a drop in the fraction of adopters $\langle s_2\rangle$ exactly at the same values of $(\theta,\lambda)$ for which the first layer experiments its transition from adopter to neutral global states, as shown in Figs. \ref{Fig:phased} and \ref{Fig:s1s2}. 

\section{Characterizing the order of transitions}
\label{sec_order}
\begin{figure}[ht]
\psfig{file=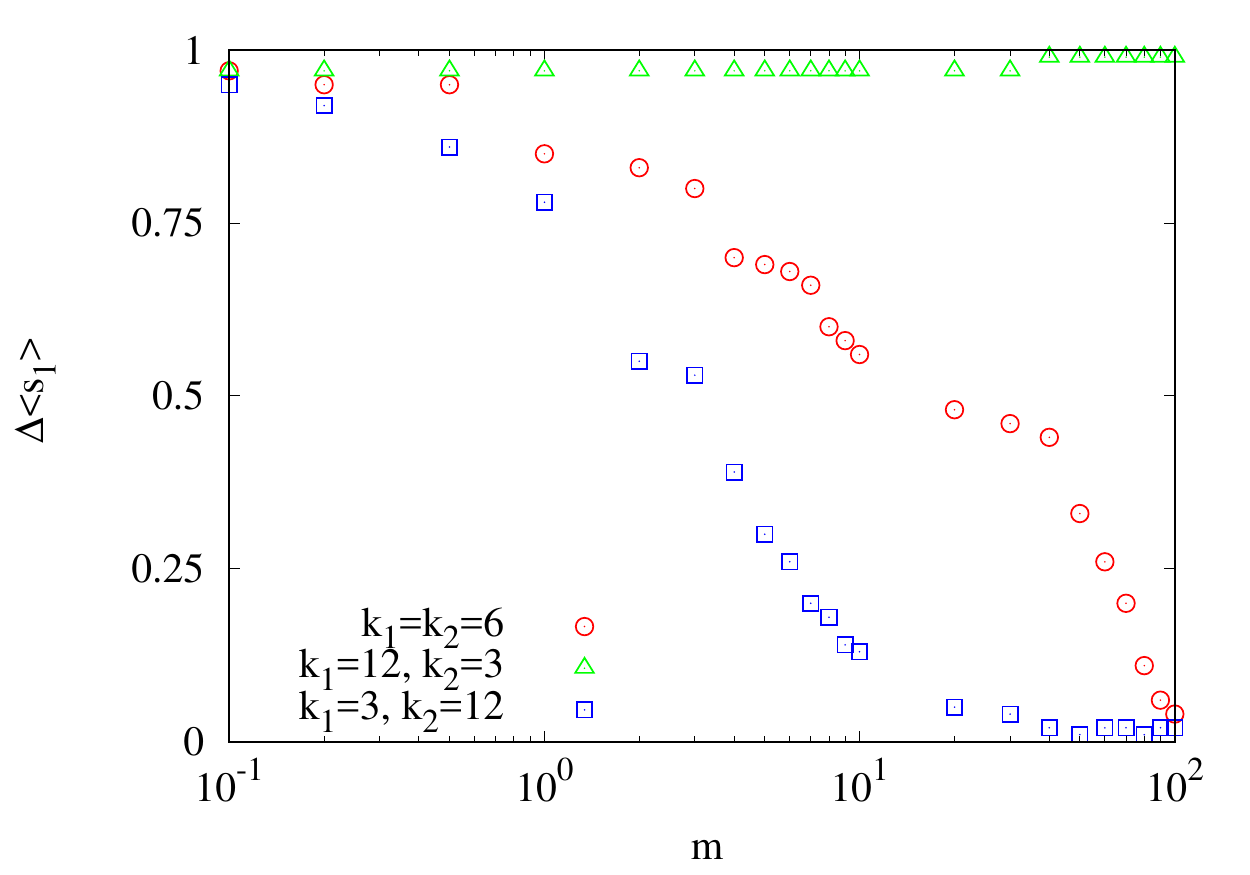,width= 1.\columnwidth, keepaspectratio = true} 
\caption{(Color online) Jump of the fraction of adopters in threshold layer $\Delta\langle s_1 \rangle$ at the transition point $\theta_c$ as a function of $m$ and for $\lambda=0.5$. Values of the $\theta_c(m)$ are shown in Fig.\ref{Fig:thetac}. Same symbols and line meanings than in figure \ref{Fig:s1s2}.}
\label{Fig:deltas}
\end{figure}

A detailed look at Fig.~\ref{Fig:s1s2} suggests that the order of the transition in the threshold layer (discontinuous in the absence of coupling to the SIS layer) might now depend on the connectivities and, most remarkably, on the exact way some transition lines are crossed. Quite generally all transitions between the different regions are discontinuous, except for the transition {\bfseries I } to {\bfseries IIb } occurring increasing $\theta$ at fixed $\lambda$ for $k_1\le k_2$ which becomes continuous for sufficiently large $m$. The change from discontinuous to continuous transition in that case is evident from Fig.~\ref{Fig:deltas} where we plot the jump of $\langle s_1\rangle$ at the transition point. Note, however, that the transition {\bfseries II } to {\bfseries I } occurring increasing $\lambda$ at fixed $\theta$ is always discontinuous. 

\begin{figure}[ht]
\centerline{\psfig{file=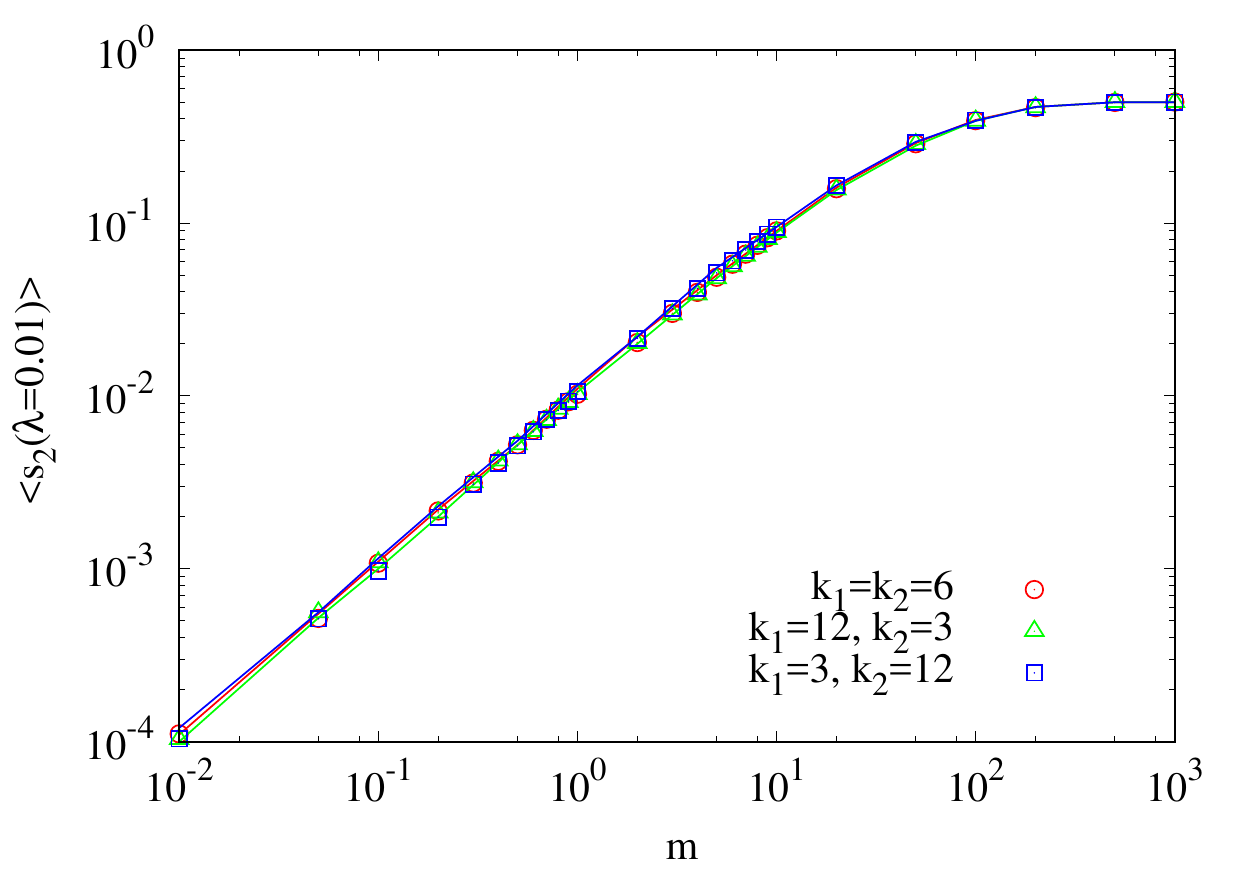,width=0.99\columnwidth}}
\caption{(Color online) Fraction of adopters right above $\lambda_c^1=0$ as a function of $m$. Different colors represent different values of $k_1$ and $k_2$. Red circles correspond to the case $k_1 = k_2$, green triangles - $k_1 > k_2$, and blue squares - $k_1 < k_2$. Symbols present the results from computer simulations for a network of $N_1+N_2=2000$ nodes, where both layers are ER random networks. Results are averaged over $500$ realizations of networks and $500$ realizations of dynamics for each network configuration. We took $\theta =\lambda$ in computer simulations. Solid lines with corresponding colors stand for analytical solutions.}
\label{Fig:s2m}
\end{figure}

When both layers are coupled the transition in the SIS layer remains continuous but moves to $\lambda=0$. We have already shown that for small number of interlayer connections $m$ and adoption probability $\lambda$ the fraction of adopters in the SIS layer grows linearly as $x_2^*\approx m\lambda$, so proving the continuous nature of the transition, see Fig. \ref{Fig:s2m} where we plot the stationary solution of equation (\ref{x1fromx2}) which fits very well the results obtained by computer simulations. Interestingly we observe also a new, second transition in the SIS layer. It appears for the same set of parameters $(\theta,\lambda)$ than a transition in the threshold layer is observed.

It appears from the numerical results that in the threshold layer the order of the transition  between the {\bfseries I } and {\bfseries IIb } regions depends, for large $m$ and $k_1\le k_2$, on whether the transition line is crossed vertically (at constant $\theta$) or horizontally (at constant $\lambda$). In fact, it changes from discontinuous to continuous when $\lambda$ is fixed and we increase $\theta$ going from phase {\bfseries I } to {\bfseries IIb } (for sufficiently large $m$ and when $k_1\le k_2$). However, our analytical calculations do not predict this change in the order of the transitions in the threshold layer. 

\begin{figure*}[ht]

\psfig{file=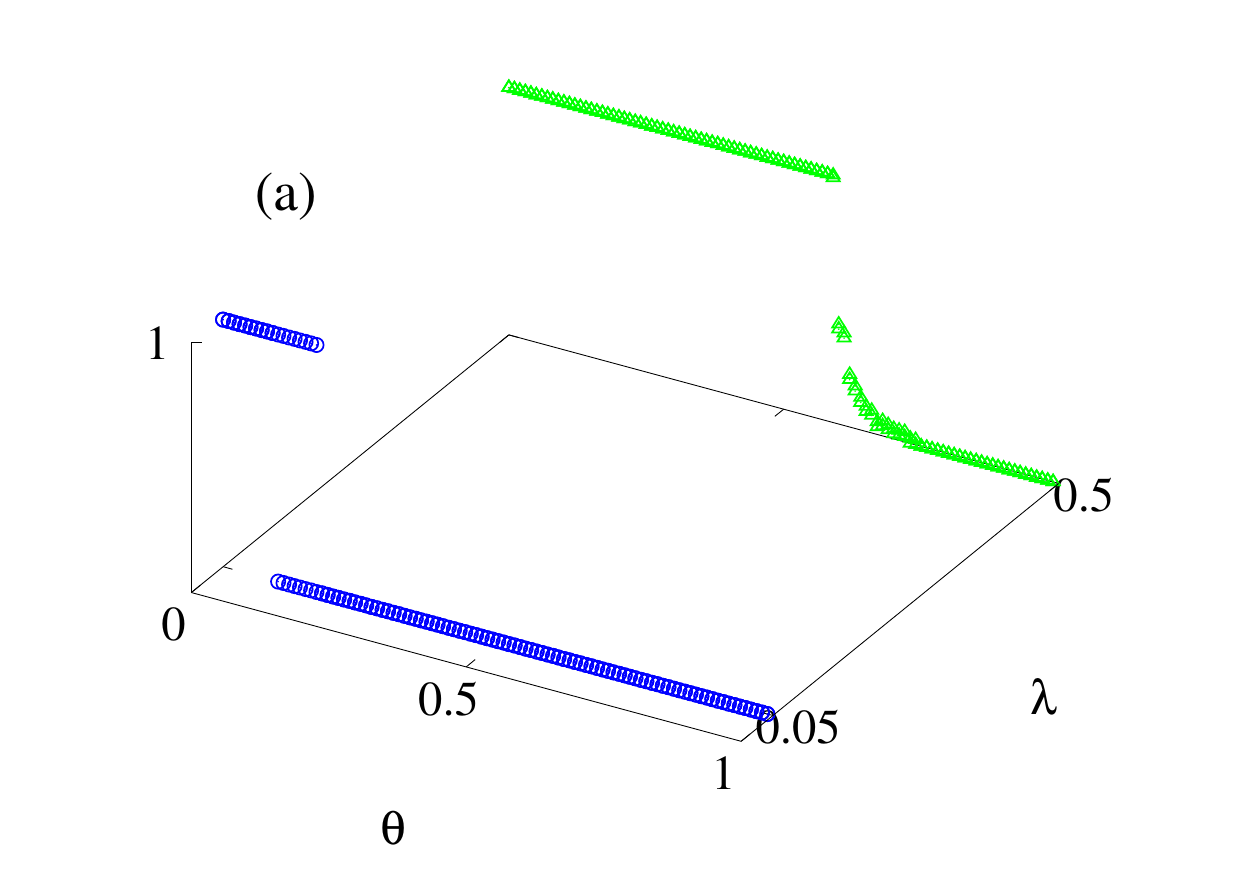,width= 0.99\columnwidth, keepaspectratio = true} \psfig{file=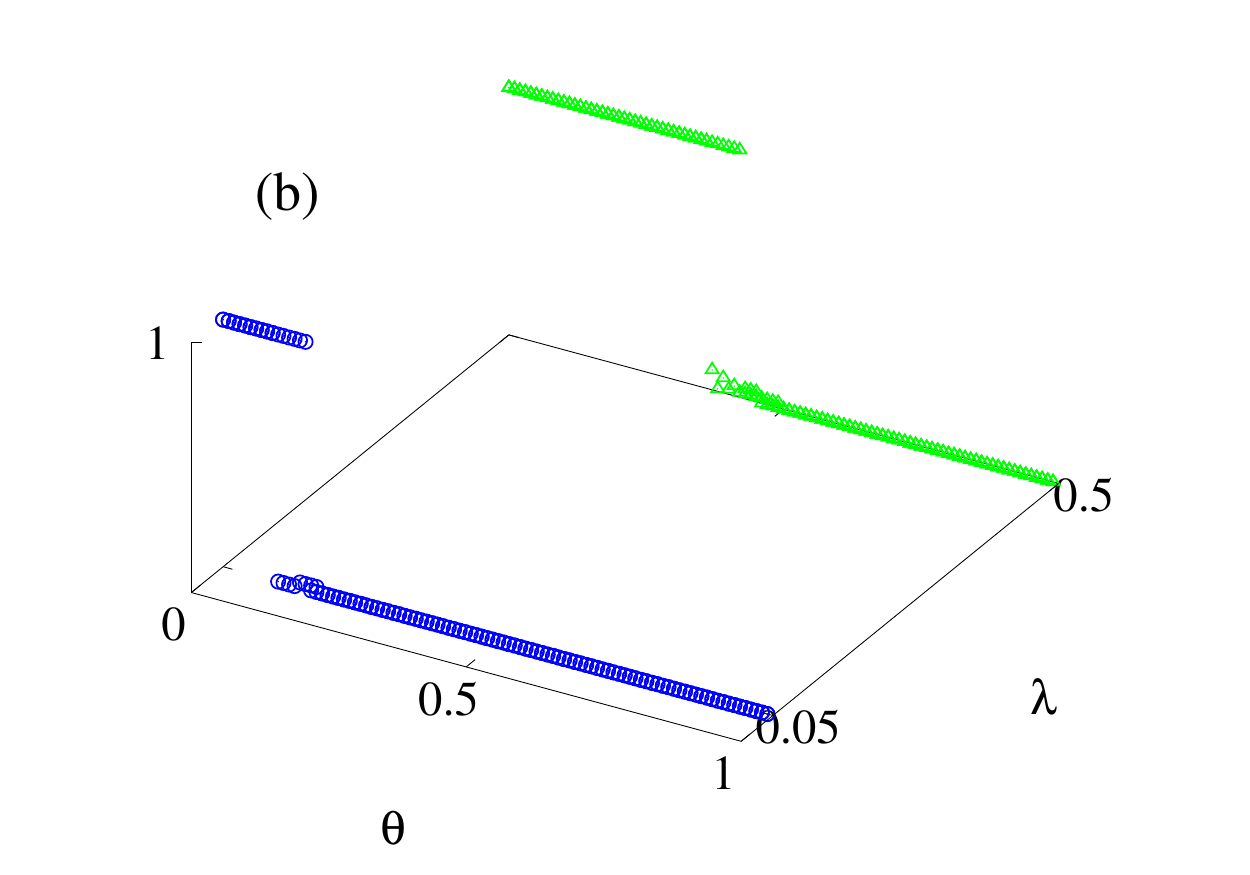,width= 0.99\columnwidth, keepaspectratio = true} \\
\psfig{file=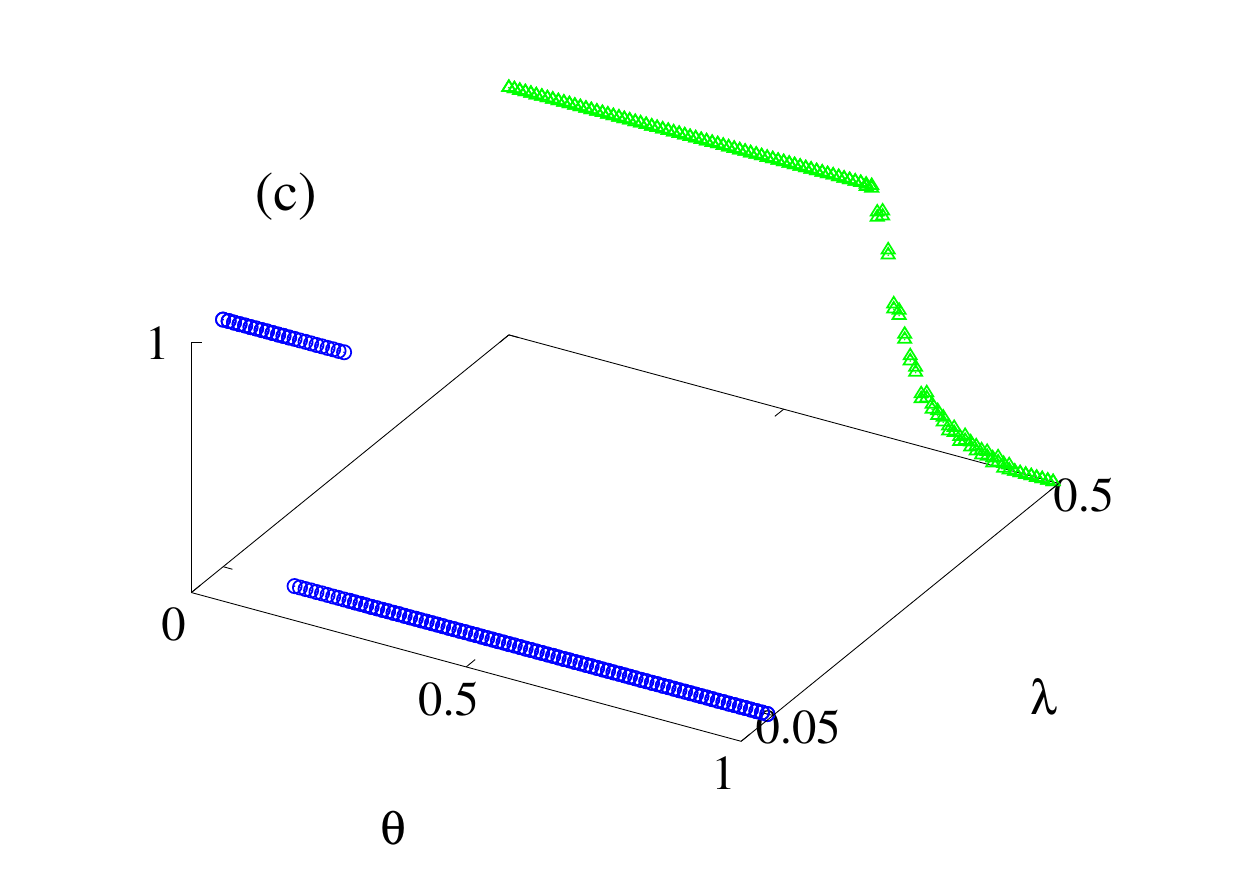,width= 0.99\columnwidth, keepaspectratio = true} \psfig{file=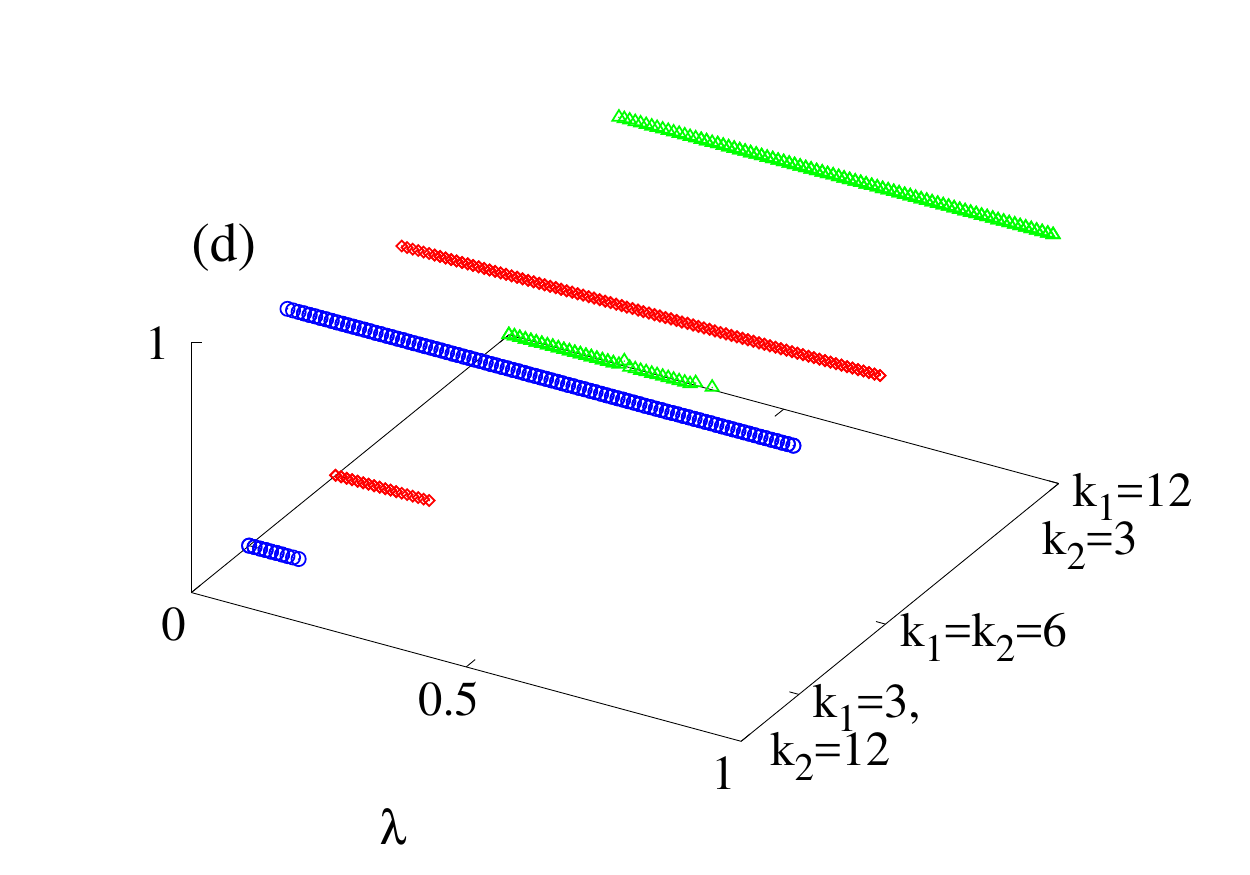,width= 0.99\columnwidth, keepaspectratio = true} 
\caption{Maxima of the fraction of adopters in threshold layer $\langle s_1\rangle_{max}$ as a function of $\theta$ --for fixed values of $\lambda$, different panels present results for different intralayer connectivities, i.e. (a) $k_1=k_2=6$, (b) $k_1=12$ and $k_2=3$, (c) $k_1=3$ and $k_2=12$-- and as a function of $\lambda$ --for fixed $\theta$, panel (d). In all cases the number of interlinks is set to $m=10$.}
\label{Fig:Ps1max}
\end{figure*}

\begin{figure}[ht]
\centerline{\psfig{file=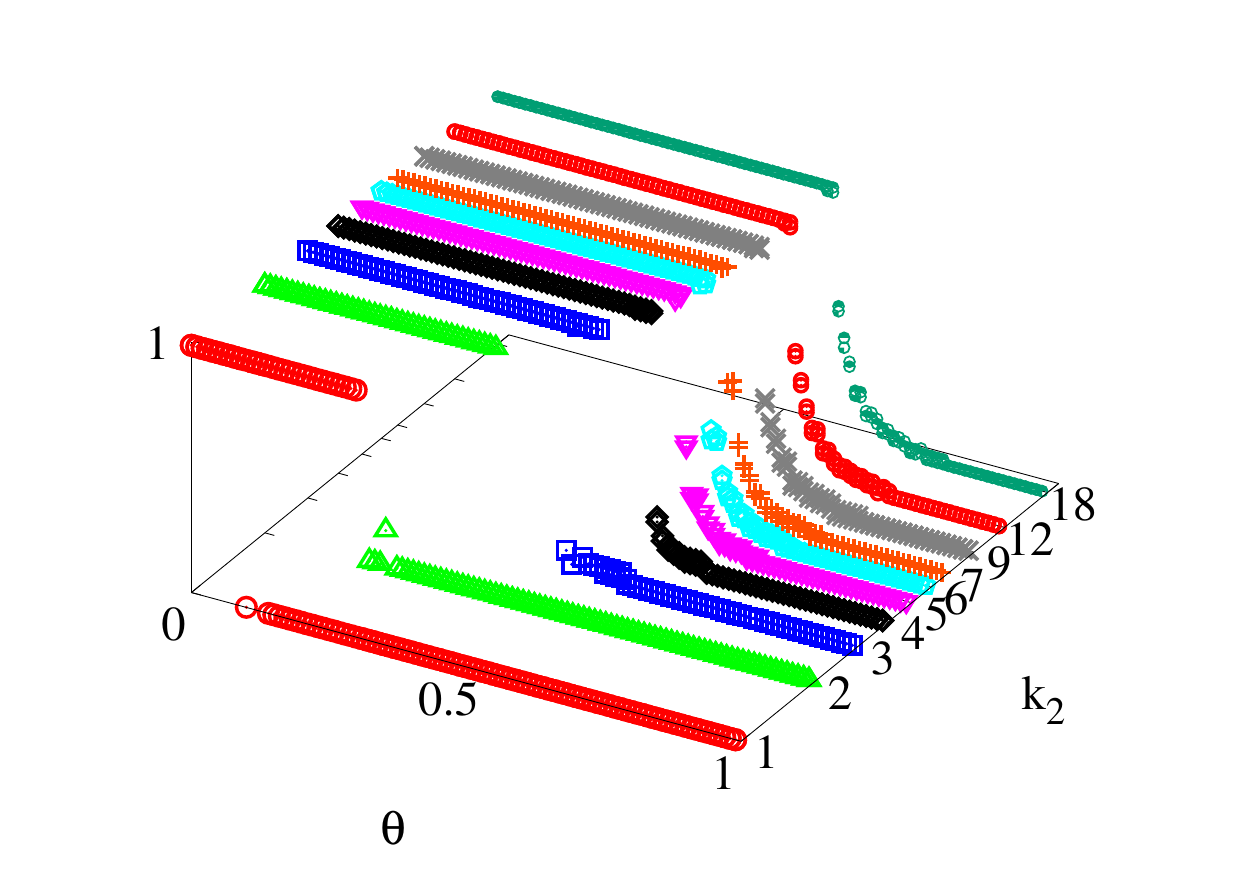,width= 0.99\columnwidth, keepaspectratio = true}}
\caption{(Color online) Maxima of the fraction of adopters in threshold layer $\langle s_1\rangle_{max}$ as a function of $\theta$ when $k_1=6$, $m=10$ and $\lambda=0.5$ for different values of $k_2$.}
\label{Fig:Ps1max_k16}
\end{figure}

To provide further numerical evidence of the order of the transitions for different connectivities $k_1$ and $k_2$ and different ways of crossing the transition lines we have studied the probability distribution $P(s_1)$ of the number of adopters in the first layer. In Figs. ~\ref{Fig:Ps1max} and \ref{Fig:Ps1max_k16} we plot the location of the maxima of this distribution for different parameter and connectivity values. Panels (a), (b), (c) in Fig. \ref{Fig:Ps1max} and Fig. \ref{Fig:Ps1max_k16} show the results when the transition line is crossed horizontally (varying $\theta$ at fix $\lambda$) while panel (d) in Fig. \ref{Fig:Ps1max} focuses on a vertical crossing (varying $\lambda$ at fix $\theta$) for different values of the inter and intralayer connectivities. 
As shown in those figures, the transition in the threshold layer crossing horizontally from {\bfseries I } to {\bfseries IIa} occurring at $\lambda<1/k_2$ remains discontinuous for all $k_1$, $k_2$ cases --blue circles in panels (a), (b) and (c)-- as so does the analogous transition from {\bfseries I } to {\bfseries IIb} occurring at $\lambda>1/k_2$ when $k_1>k_2$ --green triangles in panel (b). This discontinuous nature of the transition is clearly evidenced by the coexistence of the maxima at $s_1=0$ and $s_1=1$ for a range of values of $\theta$. In fact, in the numerical simulations, one can even observe the typical hysteresis behavior typical of a discontinuous transition. However, the same transition from {\bfseries I } to {\bfseries IIb} in the case $k_1 \leq k_2$ and for a sufficiently large number $m$ of interlink connections becomes continuous: only a single maximum of the distribution, varying continuously from $s_1=1$ to $s_1=0$ as a function of $\theta$, is observed --green triangles in panels (a) and (c). To show that our observation does not depend on the specific choice of $k_1$ and $k_2$ values we plot in Fig. \ref{Fig:Ps1max_k16} the maxima of the distribution $P(s_1)$ when $k_1=6$, $m=10$ and for different values of $k_2$. We see that the order of the transition changes from discontinuous to continuous when increasing intraconnectivity in SIS layer, $k_2$. Only when $k_1\gg k_2$ the transition remains discontinuous. For this particular example, the order of the transition changes to continuous for $k_2=4$. Fig. \ref{Fig:Ps1max} (d) shows evidence of the discontinuous character of the transition when one crosses instead the transition line vertically (at constant $\theta$) from {\bfseries I } to {\bfseries II}. As mentioned before, although the second transition in the SIS layer is caused by the transition in the threshold layer, the order of the transitions in both layers agree only when the transition line is crossed horizontally ($\lambda$ constant), but differs for vertical crossing ($\theta$ constant). In the latter case the transition in the threshold layer is discontinuous but at the same time the SIS layer experiences a continuous transition with a big slope exhibiting substantial jumps in the average $\langle s_2 \rangle$.

\section{Conclusions}
\label{sec_conclusions}
In summary, we have considered the competition between two different, simple and complex, adoption processes as a specific case of competition of a continuous and a first order transition on interdependent networks, and we have developed a mean-field approach appropriate to describe this situation. We have found that with the presence of interlayer connections the system reveals a wider range of parameters where global adoption takes place. Furthermore, both threshold and SIS layers change their behavior quantitatively and qualitatively. In the threshold layer the critical value $\theta_c$ increases with the interlayer connectivity $m$, whereas in the case of an isolated single network it would decrease with average connectivity. The transition remains discontinuous except in the case of asymmetric intralayer connectivities $k_1 \leq k_2$ and large intralayer connectivity $m$, when it becomes continuous. We also find that the critical threshold reaches a local maximum, $\theta_c>0.5$, located at intermediate values of $m$. In the SIS layer the original transition remains continuous but it moves to $\lambda_c(m)= 0$ for any $m\neq 0$, signaling the disappearance of the neutral state. A new transition in SIS layer between regions of low and large number of adopters appears caused by the interlayer coupling. This new transition can be continuous or discontinuous according to the particular values of the inter and intralayer connectivities. Remarkably the nature of the transitions in both layers might depend on the direction in which the transition lines are crossed. Our results indicate that interconnection can result in new transitions and modifications of the nature of pre-existing transitions, opening the way to further research on universal characteristics of the coupling of network transitions of different order and their dependence on inter and intralayer connectivities. 

\section*{Acknowledgments}
This work was supported by FEDER (EU) and MINECO (Spain) under Grant ESOTECOS FIS2015-63628-C2-R.

\section*{Appendix A: Mean-field approach for complex adoption}

We develop in this appendix an approach based on a mean-field type approximation in which local fractions are replaced by global averages.

Let us first introduce the general notation. We denote by $s_{1,i}(t)$, $i=1,\dots,N_1$ the states of agents in the first layer and $s_{2,i}(t)$, $i=1,\dots,N_2$ those in the second layer. An agent $i$ in layer $\ell$ is said to be in the adopter state at time $t$ if $s_{\ell,i}(t)=1$; otherwise, it is in the neutral (non-adopter) state when $s_{\ell,i}(t)=0$. We will use $\langle s_{1,i}(t)\rangle_\textrm{n}$ for the fraction of neighbors of agent $(1,i)$ who are adopters, i.e. $\langle s_{1,i}(t)\rangle_\textrm{n}=\frac1{n_{1,i}}\sum_{(\ell,j)\in n_{1,i}}s_{\ell,j}(t)$, being $n_{1,i}$ the set of neighbors of $(1,i)$ in both layers and $s_{\ell,j}$ the value of the state of such a neighbor which might belong to the first layer, $s_{1,j}$, or to the second layer, $s_{2,j}$. Once selected, the state $s_{1,i}(t)$ updates according to the following dynamical rule: if $\langle s_{1,i}(t)\rangle_\textrm{n}$ is smaller than the threshold $\theta$ nothing happens; otherwise, it becomes an adopter. This can be written as
\begin{equation}
s_{1,i}\left(t+\tau\right)=\begin{cases}
s_{1,i}\left(t\right), & \mbox{if} \; \langle s_{1,i}(t)\rangle_\textrm{n} <\theta, \\
1, & \mbox{if}\;\langle s_{1,i}(t)\rangle_\textrm{n} \geq \theta. \end{cases}
\label{thrrules_appendix}
\end{equation}
Note that, according to this rule, once a node becomes an adopter it cannot go back to the neutral state. Therefore, with the course of time the fraction of adopters in the system can either increase or stay unchanged, but never decrease.

We now aim at deriving an approximate equation for the evolution of the fraction of the number of adopters in layer 1, $s_1(t)=\frac{1}{N_1}\sum_{i=1}^{N_1}s_{1,i}(t)$. We follow closely \cite{martins2009,tessone2009} in the derivation. The ensemble average $\langle s_1(t)\rangle$ evolves according to the general, exact, relation:
\begin{equation}
N_1\langle s_1(t+\tau)\rangle=N_1\langle s_1(t)\rangle +\langle s_{1,i}(t+\tau) -s_{1,i}(t)|\{ s(t)\} \rangle,
\label{thr1_appendix}
\end{equation}
\\[5pt]where $\{ s(t)\}=\left(s_{1,1}(t),...,s_{1,N_1}(t), s_{2,1}(t),\dots,s_{2,N_2}(t)\right)$ denotes the particular realization of the state variables and $\langle \cdots|\cdots\rangle$ means a conditional average. Considering that time (as measured in Monte Caro units) increases by $\tau=1/(N_1+N_2)$ after one individual update, we can write Eq.~(\ref{thr1_appendix}) in the form
 \begin{eqnarray}
&\beta&\frac{\langle s_1(t+\tau)\rangle -\langle s_1(t)\rangle}{\tau}=\langle s_{1,i}(t+\tau) -s_{1,i}(t)|\{ s(t)\} \rangle=\notag \\ &=&-\langle s_1(t)\rangle+\langle s_{1,i}(t+\tau)|\{ s(t)\}\rangle
\label{thr2_appendix}
 \end{eqnarray}
with $\beta=\frac{N_1}{N_1+N_2}$
We now make a mean-field type approximation and consider that the fraction of neighbors which are adopters $\langle s_{1,i}(t)\rangle_\textrm{n}$ is independent of the site $i$. Hence, the probability that the fraction of adopters in the neighbourhood of the randomly chosen node $i$ is at least $\theta$ is approximated by $\text{Prob}\left[\langle s_{1,i}(t)\rangle_\textrm{n}\ge \theta\right]\approx\text{Prob}\left[\langle s_{1}(t)\rangle_\textrm{n}\ge \theta\right]$, being $\langle s_{1}(t)\rangle_\textrm{n}$ the average value of $\langle s_{1,i}(t)\rangle_\textrm{n}$ over all sites $i=1,\dots,N_1$. Using the dynamical rules described in Eq.~(\ref{thrrules_appendix}) we derive:
\begin{eqnarray}
\langle s_{1,i}&(t+\tau)&|\{ s(t)\}\rangle=\left(1-\text{Prob}\left[\langle s_{1}(t)\rangle_\textrm{n}\ge \theta\right]\right)\times\langle s_{1}(t)\rangle \notag \\ &&\qquad\qquad +\text{Prob}\left[\langle s_{1}(t)\rangle_\textrm{n}\ge \theta\right]\times 1 \notag\\
&=&\langle s_{1}(t)\rangle+(1-\langle s_{1}(t)\rangle)\text{Prob}\left[\langle s_{1}(t)\rangle_\textrm{n}\ge \theta\right].
\end{eqnarray}

Replacing in Eq.~(\ref{thr2_appendix}) and treating the left hand side as a time derivative we obtain
\begin{equation}
\beta\dfrac{d\langle s_1(t)\rangle}{dt}= \left(1-\langle s_1(t)\rangle\right)\text{Prob}\left[\langle s_{1}(t)\rangle_\textrm{n}\ge\theta\right],\label{thr3_appendix}
\end{equation}
which is Eq.(\ref{g1}) in the main text.

\section*{Appendix B: Mean-field approach for simple adoption}
We recall that the rules of the adoption process in the SIS layer are the following: at time $t$ an agent from layer 2 is randomly selected, let $s_{2,i}(t)$ be the state of this agent. If $s_{2,i}(t)=1$ (adopter) it goes back to the neutral state $s_{2,i}(t+\tau)=0$. If $s_{2,i}(t)=0$ (neutral) then it visits sequentially all its neighbors, having a probability $\lambda$ of becoming an adopter from the interaction with anyone of them (of course, if it becomes adopter in a given interaction, it is not necessary to continue the sequence of interactions with the neighbors). Namely,
\begin{equation}
s_{2,i}\left(t+\tau\right)=\begin{cases}
1, & \mbox{if} \; s_{2,i}(t)=0\; \mbox{and adoption from} \\
 & \mbox{any neighbour happens,} \\
0,& \mbox{if} \; s_{2,i}(t)=0\; \mbox{and adoption does}\\
 & \mbox{ not happen,} \\
0, & \mbox{if}\; s_{2,i}(t)=1. \end{cases}
\label{sisrules_appendix}
\end{equation}

By a similar reasoning to the one developed before for the threshold layer, we can derive an exact evolution equation for the ensemble average of the fraction of adopters in the SIS layer $s_2(t)=\frac{1}{N_2}\sum_{i=1}^{N_2}s_{2,i}(t)$
\begin{widetext}
\begin{eqnarray}
(1-\beta)\frac{\langle s_2(t+\tau)\rangle -\langle s_2(t)\rangle}{\tau}= \langle s_{2,i}(t+\tau) -s_{2,i}(t)|\{ s_2(t)\} \rangle=-\langle s_2(t)\rangle+\langle s_{2,i}(t+\tau)|\{ s(t)\}\rangle
\label{sis2_appendix}
\end{eqnarray}
\end{widetext}
According to the dynamical rules Eq.~(\ref{sisrules_appendix}), the conditional average is
\begin{eqnarray}
\langle s_{2,i}(t+\tau)|\{ s(t)\}\rangle=(1-\langle s_{2,i}\rangle)\times \textrm{Prob[A]},
\end{eqnarray}
where Prob[A]=Prob[Adoption occurs from any neighbour]=$1-$Prob[Adoption does not occur from any neighbour].
If $\kappa_i$ is the number of adopter neighbours in any layer of site $(2,i)$ the probability that adoption does not occur for that site is $(1-\lambda)^{\kappa_i}$. In the mean-field approximation we will replace this probability by the average probability $\langle (1-\lambda)^{\kappa_i}\rangle$ over all nodes. We will further assume that the number of adjacent adopters is given by a Poisson distribution (as in ER networks), $P(\kappa)=\frac{\langle\kappa\rangle^{\kappa}e^{-\langle\kappa\rangle}}{\kappa !}$, leading to
\begin{equation}
\langle (1-\lambda)^{\kappa_i}\rangle=\sum_{\kappa=0}^{\infty}\frac{\langle\kappa\rangle^{\kappa}e^{-\langle\kappa\rangle}}{\kappa !}(1-\lambda)^\kappa=e^{-\lambda\langle \kappa\rangle}.
\end{equation}
We replace $\langle \kappa\rangle=k_2 \langle s_2 \rangle + m \langle s_1\rangle$. Thus probability that adoption happens is
\begin{equation}
\textrm{Prob[Adoption]}=1-e^{-\lambda\left(k_2 \langle s_2 \rangle + m \langle s_1\rangle\right)}.
\label{g2_appendix}
\end{equation}
Replacing in Eq.~(\ref{sis2_appendix}) and identifying the left side as a time derivative we obtain the mean-field equation for the fraction of adopters in the SIS layer
 \begin{equation}
(1-\beta)\dfrac{d\langle s_2(t)\rangle}{dt}= -\langle s_2\rangle+\left(1-\langle s_2\rangle\right)\left(1-e^{-\lambda\left(k_2 \langle s_2 \rangle + m \langle s_1\rangle\right)}\right),
\label{sis4_appendix}
\end{equation}
which is Eq.(\ref{sis4}) in the main text.

\end{document}